\documentclass[fleqn,12pt,twoside]{article}
\usepackage[headings]{espcrc1}

\readRCS
$Id: espcrc1.tex,v 1.2 2004/02/24 11:22:11 spepping Exp $
\usepackage{graphicx}
\usepackage{psfig}
\usepackage[figuresright]{rotating}

\newcommand{\AmS}{{\protect\the\textfont2
  A\kern-.1667em\lower.5ex\hbox{M}\kern-.125emS}}

\title{Lattice QCD at finite temperature}

\author{P. Petreczky\address[MCSD]{
        Physics Department and RIKEN-BNL, \\
        Brookhaven National Laboratory, Upton NY 11973, USA}%
        \thanks{This work has been supported by  
        U.S. Department of Energy under contract DE-AC02-98CH10886.}
       }

\runtitle{1-column format camera-ready paper in \LaTeX}
\runauthor{P. Petreczky}

\begin{document}

\maketitle

\begin{abstract}
I discuss recent developments in finite temperature lattice QCD,
including the calculation of the transition temperature, equation of
state, color screening and meson spectral functions.
\end{abstract}

\section{Introduction}

In recent years considerable progress has been made in understanding QCD 
at finite temperature using lattice simulations. Calculations
of bulk thermodynamic quantities have been done in the realistic case 
of 2+1 flavors with sufficiently small quark masses using different versions of improved 
staggered actions \cite{milc04,hyp,milclat05,fodor05,us06}. This reduces or can even
eliminate the uncertainty of extrapolations in the quark mass in
different thermodynamics observables. 

One of the most prominent features of the deconfined phase of QCD
is color screening. In non-Abelian gauge theories it is a very
complicated phenomenon. Recently some progress in understanding
color screening has been made by calculating the free energy of
static charges on the lattice \cite{okacz02,digal03,okaczlat03,petrov04}.

For many years lattice QCD did not provide any information on the
properties of real time excitations at finite temperature. The situation
has changed since few years and first calculations of meson 
spectral functions at finite temperature appeared \cite{karsch02,umeda02,asakawa04,datta02,datta04,jhw}.
In this contribution I am going to review some of these 
recent developments.

\section{Transition temperature and the equation of state}

One of the most interesting question for the lattice 
is the question about the nature of the finite temperature transition 
and the value of the temperature $T_c$ where it takes place. 
For very heavy quarks we have a 1st order deconfining transition.
In the
case of QCD with three degenerate flavors of quarks we expect a 1st order
chiral transition for sufficiently small quark masses.
In other cases there is no true phase transition but just a rapid
crossover. 
Lattice simulations of 3 flavor QCD with improved staggered quarks (p4) using
$N_t=4$ lattices indicate that
the transition is first order only for very small quark masses, 
corresponding to pseudo-scalar meson masses of about $60$ MeV 
\cite{karschlat03}.
A recent study of the transition using effective models
of QCD  resulted in a similar estimate for the boundary in the quark mass
plane, where the transition is 1st order \cite{szepzs}.
This makes it unlikely that for the interesting case of one heavier strange
quark and two light $u,d$ quarks corresponding to $140$ MeV pion the
transition is 1st order. However, calculations with unimproved staggered
quarks suggest that the transition is 1st order for pseudo-scalar
meson mass of about $300$ MeV \cite{norman}. 
Thus the effect of the improvement is
significant and we may expect that the improvement of flavor symmetry,
which is broken in the staggered formulation, is very important.
But even when using improved staggered fermions it is necessary to do the calculations at several
lattice spacings in order to establish the continuum limit. 
Recently,  extensive calculations have been done to clarify the nature 
of the transition in the 2+1 flavor QCD for physical quark masses using
$N_t=4,~6,~8$ and $10$ lattices. 
These calculations were done using the
so-called $stout$ improved staggered fermion formulations which is even
superior to other more commonly used improved staggered actions ($p4,~asqtad$)
in terms of  improvement of flavor symmetry.  The result of this study was 
that the transition is not a true phase transition but only a rapid
crossover \cite{nature}.

Even-though there is no true phase transition in  QCD thermodynamic observables change rapidly in a small 
temperature interval and the value of
the transition temperature plays an important role. 
The flavor and quark mass dependence of 
many thermodynamic quantities is largely determined by the flavor and 
quark mass dependence of $T_c$. For example, the pressure normalized by
its ideal gas value for pure gauge theory, 2 flavor, 2+1 flavor and 3 flavor
QCD shows almost universal behavior as function of $T/T_c$ \cite{cargese}.

The current status of lattice calculations of the transition temperature
is summarized in Fig. 1a. 
Early calculations with $p4$ action have been done for 2- and 3- flavor QCD
for pion masses $m_{\pi} \ge 400$ MeV \cite{karsch01}. 
Calculations with Wilson fermions have
large uncertainties from chiral extrapolation. This is 
because the actual simulations are done in the
region  of large quark masses, corresponding to pion masses  
$m_{\pi} \ge 700 MeV$  \cite{cppacsnf2,nakamuralat05}.
There is also a calculation of $T_c$ using standard staggered fermion for the
physical  pion mass \cite{fodor04muc}.

To better understand the systematics of the determination of $T_c$ in Fig. 1b I 
show the transition temperature in units of the Sommer scale $r_0$ as
the function of the pion mass based on calculations on lattices with temporal
extent $N_t=4$ and $N_t=6$ with $p4$ action \cite{us06}.
\begin{figure}
\centerline{\hfill a  \hspace*{9cm} b \hfill}
\includegraphics[width=8cm]{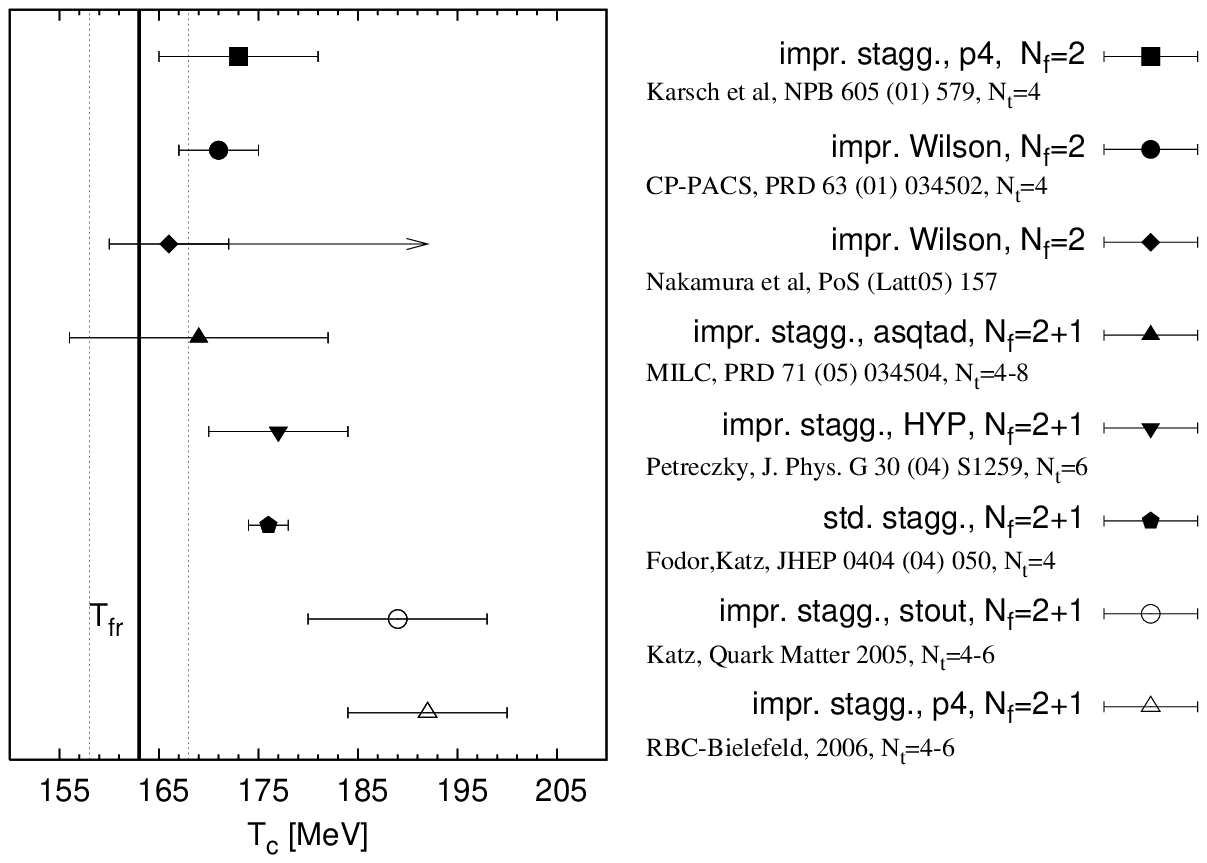}
\includegraphics[width=8cm]{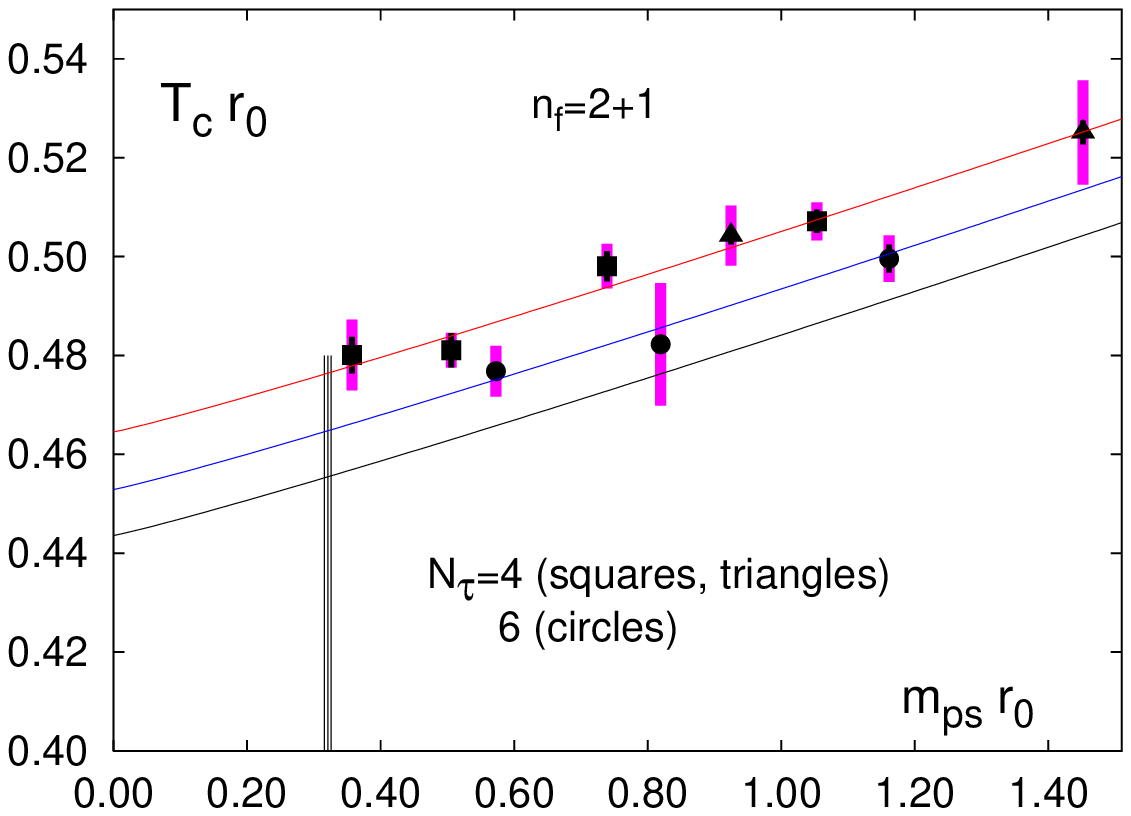}
\vspace*{-1.5cm}
\caption{Lattice data on the transition temperature from 
Refs. \cite{karsch01,milc04,hyp,us06,cppacsnf2,nakamuralat05} and the 
chemical freezout temperature at RHIC \cite{starwhite} (left). 
The transition temperature in units of the $r_0 T_c$ from Ref. \cite{us06}
as function of the pion mass.}
\vspace*{-0.4cm}
\end{figure}
The pion masses in these calculations
are smaller than 500 MeV with the smallest pion mass being around the physical
value. Note that the value of $T_c$ calculated at two different lattice 
spacings are clearly different. The thin errorbars in Fig. 1b represent the
error in the determination of the lattice spacing $a$, i.e. 
the error in $r_0/a$. There is also an error in the determination of the
gauge coupling constant $\beta_c=6/g^2$. The combined error is shown in 
Fig. 1b as a thick errorbar. For $N_t=4$ calculations the error is dominated 
by the error in lattice spacing, while for $N_t=6$ it is dominated by the
error in $\beta_c$. With the data on $r_0 T_c$ a chiral and continuum
extrapolation has been attempted using the most simple Ansatz 
$r_0 T_c(m_{\pi},N_t)=r_0 T_c|_{cont}^{chiral}+ A (r_0 m_{\pi})^d + B/N_t^2$. 
From this extrapolation on gets the continuum value $T_c r_0=0.457(7)[+8][-2]$ 
for the physical pion mass $m_{\pi} r_0=0.321$ {\cite{us06}. 
The central value was obtained using 
$d=1.08$ expected
from $O(4)$ scaling. To test the sensitivity to the chiral extrapolations
$d=2$ and $1$ have also been used. 
The resulting uncertainty is shown
as second and third error in square brackets.     
Using the
best know value of $r_0=0.469(7)$ we obtain  
$T_c=192(7)(4)MeV$ 
which is higher than the most
of the previous values. It is also significantly higher than the 
chemical freezout temperature at RHIC \cite{starwhite} which is shown in Fig.1a as the
vertical band.  
\begin{figure}
\centerline{ a  \hspace*{7cm} b }
\includegraphics[width=8cm]{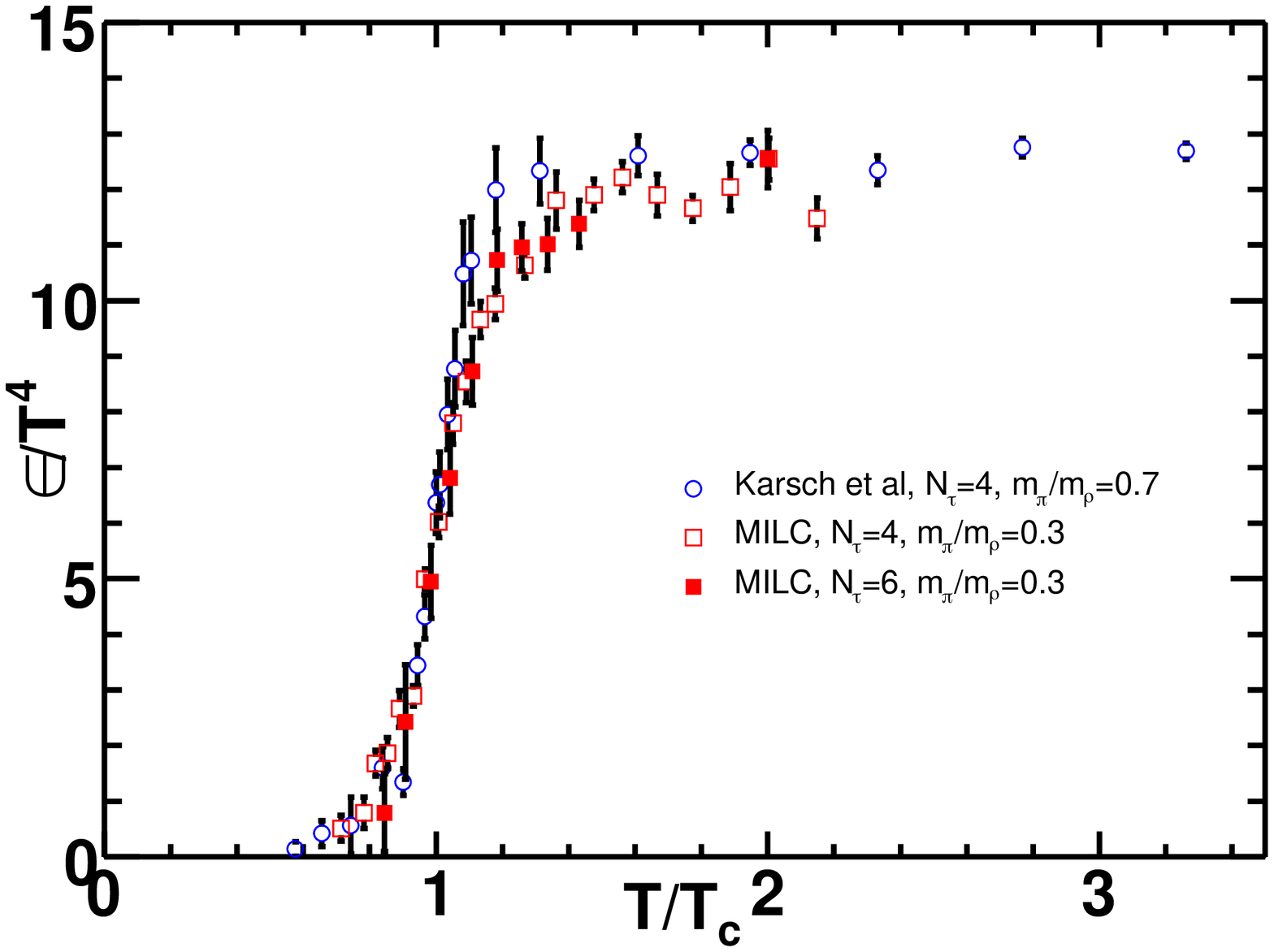}
\includegraphics[width=8.5cm]{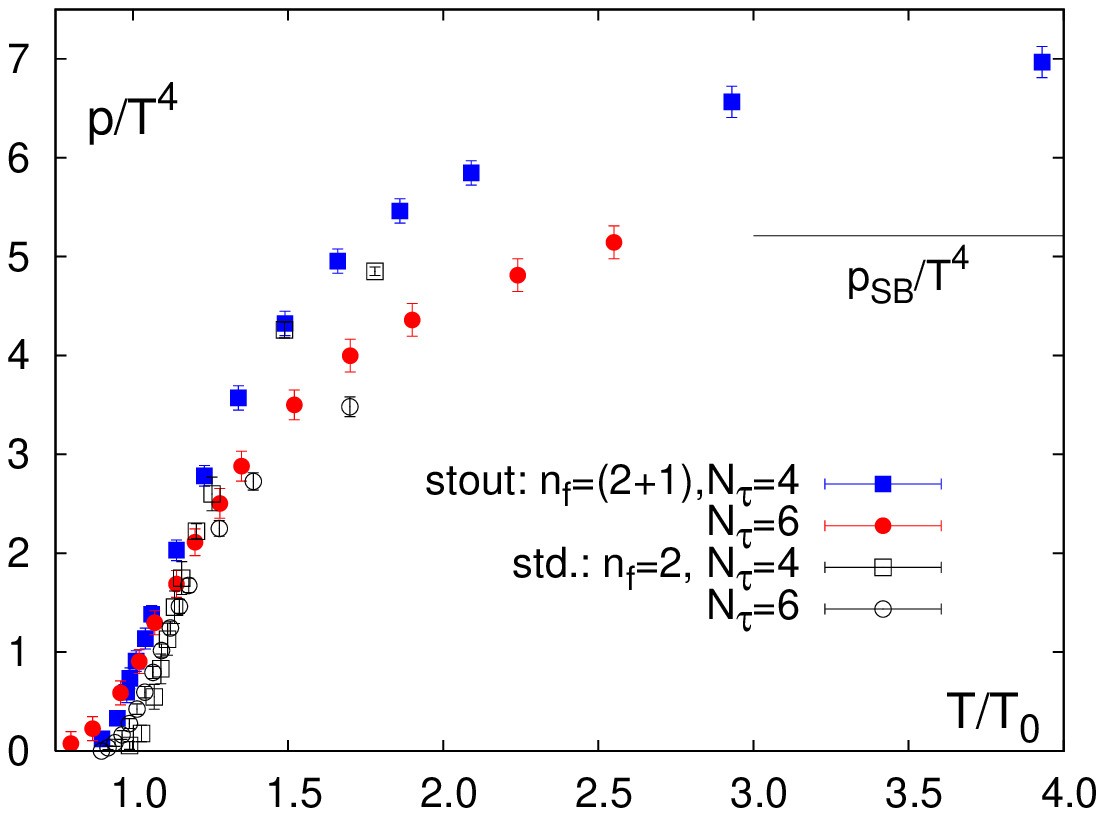}
\vspace*{-1.0cm}
\caption{The energy density
calculated with $p4$ and $asqtad$ actions (left) and the  
pressure calculated with {\em stout} action. Calculation of the
pressure with  standard staggered action is also shown.  
}
\vspace*{-0.4cm}
\end{figure}

The equation of state has been calculated using both improved action and standard staggered
fermion actions for different quark masses. The  first calculation of the equation of state
with improved action ($p4$) was done for $N_t=4$ and at quite heavy quark masses,
corresponding to pion mass of about $700$ MeV at the transition temperature \cite{karsch00}. 
The quark mass was kept fixed in units of the temperature and not in physical units \cite{karsch00}. 
These may have an effect on the high temperature limit of the pressure and energy density \cite{fodor04}.
More recent calculations with $asqtad$ and $stout$ improved action were done at smaller quark masses, namely
for pions of about $200$ and $300$ MeV for $asqtad$ and for physical pion mass for $stout$ action. 
The result of these calculations is shown in Fig. 2. In Fig. 2a I show the energy density as function
of $T/T_c$ for p4 and asqtad action.
Calculations done at $N_t=4$ and $N_t=6$ agree well with each other.
One can also see that the energy density calculated with the two 
different action agrees well despite the fact that the quark masses are significantly different. 
Close to the transition we may argue that the mass dependence of the energy density is taken care
of by the mass dependnce of $T_c$ as discussed above. 

The pressure  calculated
with the $stout$ action  is shown in Fig 2b. Unlike $p4$ and $asqtad$ the $stout$ action impoves the flavor symmetry breaking but not
the breaking of rotational symmetry which is important in the high temperature limit. 
This is probably the reason why the $N_t=4$ and $N_t=6$ results differ significantly for
temperatures $T>1.3T_c$.  In fact the cut-off dependence of the pressure for stout action in the high temperature
phase seems to be as large as the cut-off dependence in the case of standard (unimproved) staggered action \cite{milc_std_eos}.
In Ref. \cite{fodor05} it has been argued that the cutoff dependence can be taken care of by normalizing the
pressure by the ideal gass value  calculated on the finite lattice $p_{SB}^{latt}$. Indeed, $p/p_{SB}^{latt}$
shows significantly reduced cut-off dependence. However, calculations in pure gauge theory show that 
normalizing by the lattice ideal gas value may overestimate the cut-off effects \cite{boyd96}. Therefore future studies 
are needed to settle this issue.

\section{Static quark and temporal meson correlators}
To study screening of static charges in high temperature QCD it is
customary to calculate the color singlet $F_1$,  octet $F_8$ and  averaged free energies of
a static quark anti-quark pair \cite{mclerran,nadkarni}
\begin{eqnarray}
&&
e^{-F_1(r,T)+C}=
\frac{1}{3} \langle {\rm Tr} W(\vec{r}) W^{\dagger} (0) \rangle ,\\
&&
e^{-F(r,T)+C} = \frac{1}{9} \langle 
{\rm Tr} W(\vec{r}) {\rm Tr} W^{\dagger} (0) \rangle 
=\frac{1}{9} \exp(-F_1(r,T))+\frac{8}{9} \exp(-F_8(r,T)).
\label{freee}
\end{eqnarray}
While the color averaged free energy is gauge invariant, the singlet and octet
free energies need gauge fixing. Usually the Coulomb gauge is used 
\cite{okacz02,digal03,okaczlat03,petrov04,okacz04,okacz05}. There are also gauge invariant
definitions of the singlet free energy in terms of Wilson loop \cite{nadkarni,owe04} and eigenfunctions
of the covariant Laplacian \cite{owe02}.
In the high temperature leading order
perturbation theory all definition give the same result. Also in the zero temperature
limit different definitions should agree \cite{owe02}.   
The normalization constant $C$ should be fixed such that the free energy of static
quark anti-quark pair coincides with the zero temperature potential \cite{okacz02}.
In Fig.3a I show the results of the color singlet free energy in Coulomb gauge in 3
flavor QCD \cite{petrov04}. As one can see at very short distances the free energy is temperature independent
and agrees with the zero temperature potential. It approaches a constant, $F_{\infty}$, at large
separation for all temperatures. However, as temperature is increasing $F_{\infty}$ decreases 
as well as the distance where the free energy effectively flattens off.
This can be thought of as  manifestation of color screening.
Similar
results have been obtained in pure gauge theory \cite{okacz02,digal03,okacz04} 
and two flavor QCD \cite{okacz05} with the obvious difference that in pure gauge theory 
$F_{\infty}$ is infinite, as the free energy rises linearly with separation below
the deconfinement temperature \cite{okaczlat03}.  
One can also calculate the entropy and the internal energy of the two static
quarks, $S_{\infty}=-\partial F_{\infty}/\partial T$, $U_{\infty}= F_{\infty}+T S_{\infty}$.
In Fig. 3b I show 
the entropy of two separated static quarks as function of the temperature for 2 and 3 flavor
QCD as well as pure gauge theory. There is a large peak in  $S_{\infty}$ at the transition 
temperature. At large  temperature $S_{\infty}$ seems to approach a constant value. The internal energy is
very close to zero at high temperatures.
\begin{figure}
\centerline{ a  \hspace*{7cm} b }
\hspace*{-0.4cm}\includegraphics[width=8.2cm]{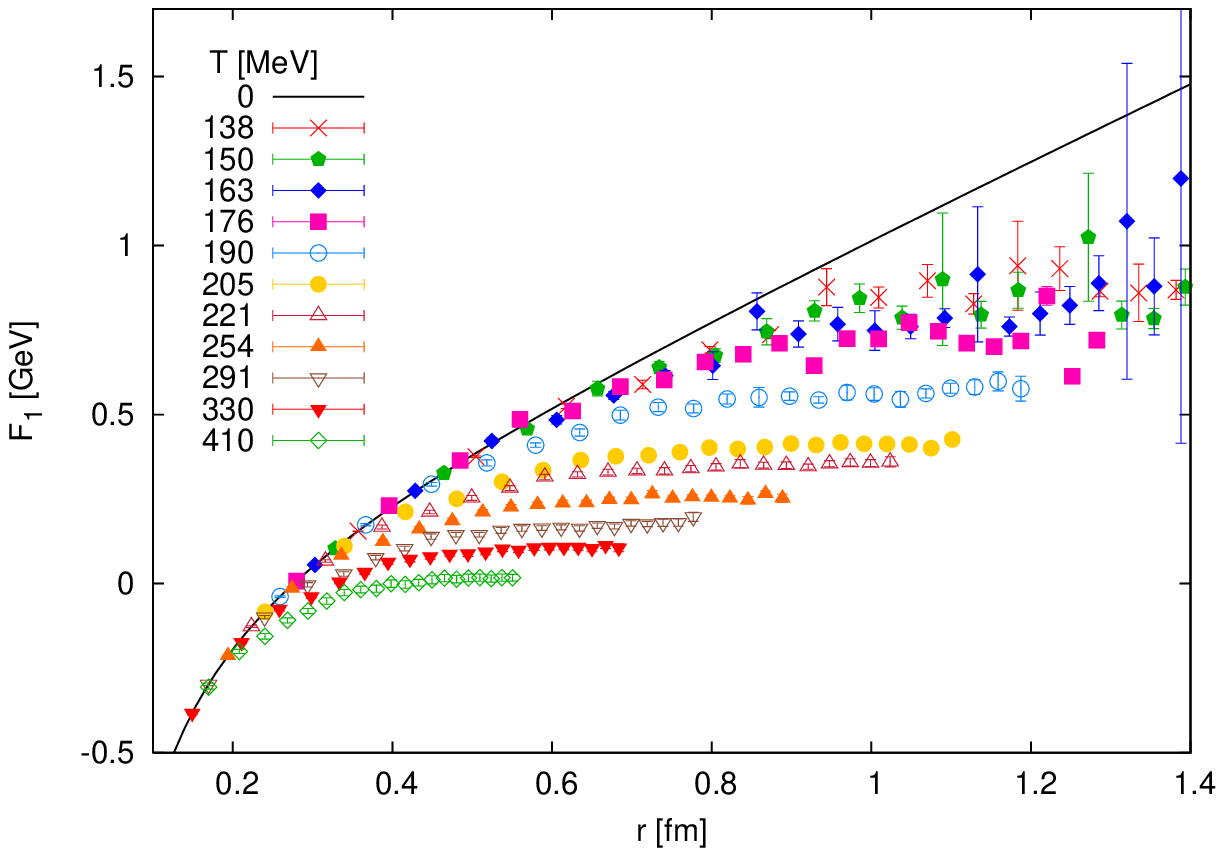}
\includegraphics[width=8.6cm]{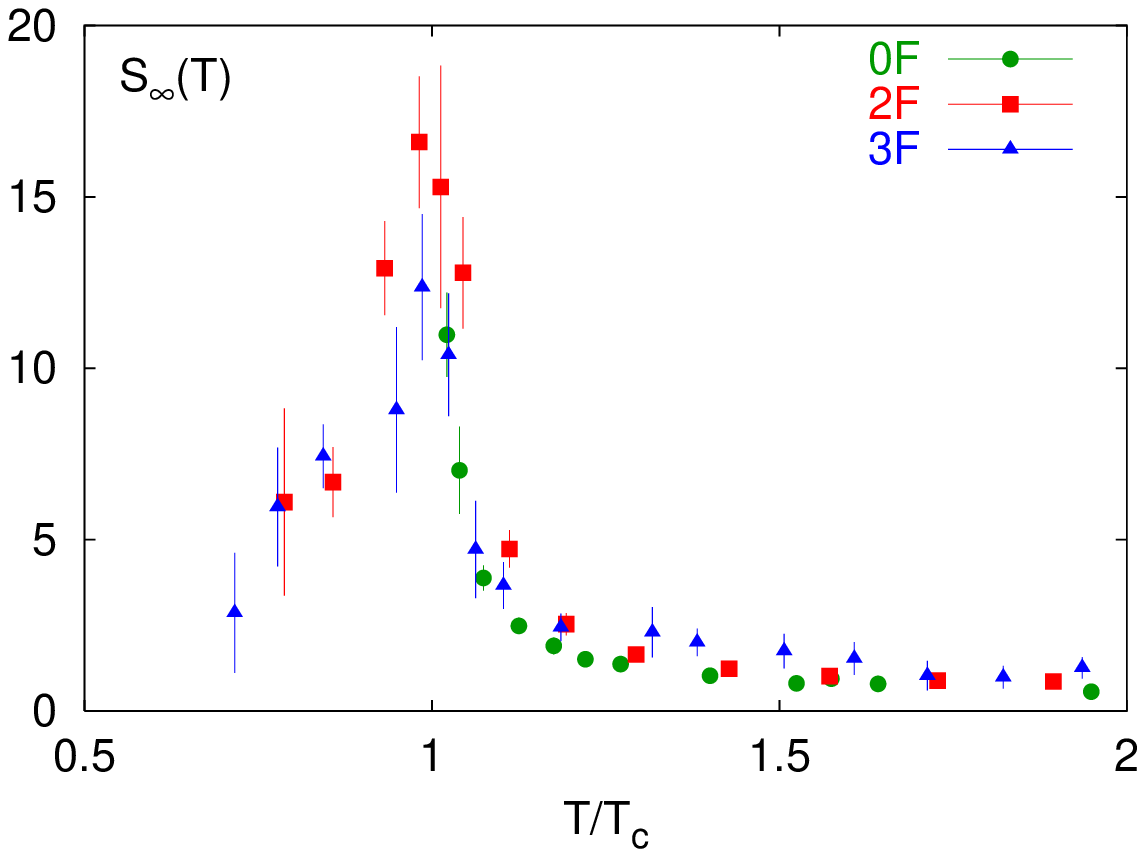}
\vspace*{-1.1cm}
\caption{The singlet free energy in 3 flavor QCD \cite{petrov04} at
different temperaure (left) and the entropy of isolated quark and anti-quark
in pure gauge theory, 3- and 2-flavor QCD \cite{mehard04,okacz05} (right). }
\vspace*{-0.4cm}
\end{figure}
The decrease
of $F_{\infty}$ at high temperature is due to the presence of the entropy contribution, $-T S_{\infty}$, which at
asymptotically high temperatures is given by $- \frac{4}{3} T \alpha_s (m_D/T) \sim g^3 T$ and thus
dominates the free energy \cite{mehard04}.  Here $m_D$ is the leading order Debye mass. 
The internal energy is zero at order ${\cal O}(g^3)$ and receives contributions from higher order.
The above relation can be used for a non-perturbative
definition of the Debye mass. 

The above analysis, done for the singlet free energy, can also be done for the color averaged
free energy in the same way. At very small distances we have $F_1=-(4/3) \alpha_s/r$ and
$F_8=+(1/6) \alpha_s/r$ and thus $F=F_1+ T \ln 9$ \cite{okacz02}.  The color averaged free energy approaches
the same constant  $F_{\infty}$, thus this quantity is gauge independent. However, the temperature dependence
of $F$ is much stronger than of $F_1$ and one should consider much smaller distances to normalize
it properly. This makes the lattice study of this quantity more difficult \cite{okacz02}. This is the main reason why most 
of the recent lattice calculations concentrated on the singlet free energy.

Based on the simple picture of color screening it has been argued by Matsui and Satz that quarkonia should
disappear at temperatures slightly above the deconfinement temperature \cite{MS86}. To verify this conjecture
and gain information about the quarkonium properties above the deconfinement transition meson
spectral functions have to be calculated. On the lattice this can be done using the relation between  Euclidean
and real time meson correlators, which in terms of the spectral function can be written as :
\begin{equation}
G(\tau, T) = \int_0^{\infty} d \omega
\sigma(\omega,T) K(\tau,\omega,T) ,~~
K(\tau,\omega,T) = \frac{\cosh(\omega(\tau-1/2
T))}{\sinh(\omega/2 T)}.
\label{eq.kernel}
\end{equation} 
Given the data on the Euclidean meson correlator $G(\tau, T)$ the meson spectral function can be calculated 
using the Maximum Entropy Method \cite{mem}. For charmonium this was done by using correlators calculated on
isotropic lattices \cite{datta02,datta04} as well as  anisotropic lattices \cite{umeda02,asakawa04,jhw}.  Different charmonium
states are identified as peaks in the spectral functions. It was shown that the spectral functions for the ground state
charmonia change little across the deconfinement transition \cite{umeda02,asakawa04,datta04,jhw}. This suggests that
$1S$ ($J/\psi$, $\eta_c$) charmonia can survive till temperatures as high as $1.6T_c$.  
Spectral functions of excited $1P$ ($\chi_c$) states show large changes already at $1.1T_c$ \cite{datta02,datta04,jhw}.
This is shown in Fig. 4 where the spectral functions in the pseudo-scalar and scalar channels 
calculated on anisotropic lattice \cite{jhw} are shown.
We see no change in the pseudo-scalar spectral function which corresponds to $\eta_c$ state. On the other
hand the scalar spectral function corresponding to $\chi_c$ state shows dramatic modifications. 
Note that the spectral functions in Fig. 4 are zero above some energy. This is due finite lattice spacing which
limits the maximal energy $\omega$. 
\begin{figure}
\centerline{ a  \hspace*{7cm} b }
\hspace*{-0.4cm}\includegraphics[width=8.2cm]{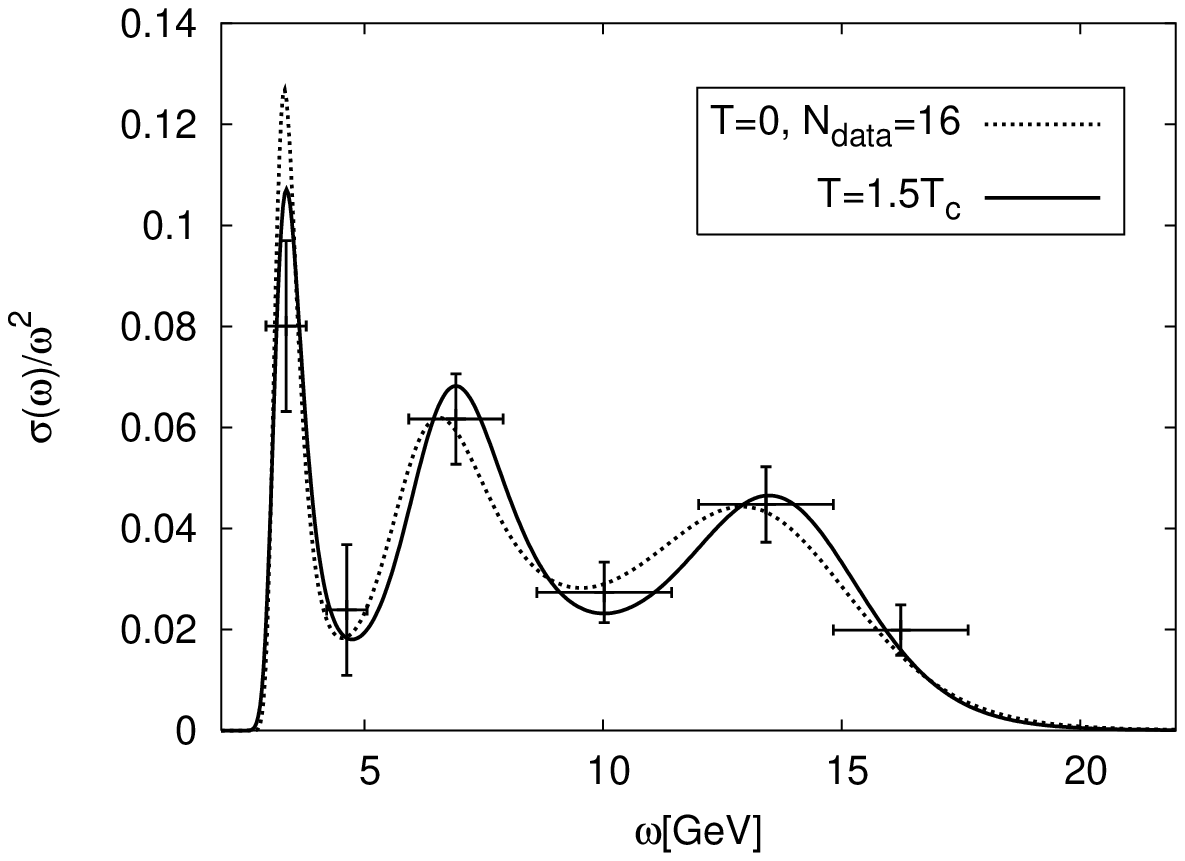}
\includegraphics[width=8.6cm]{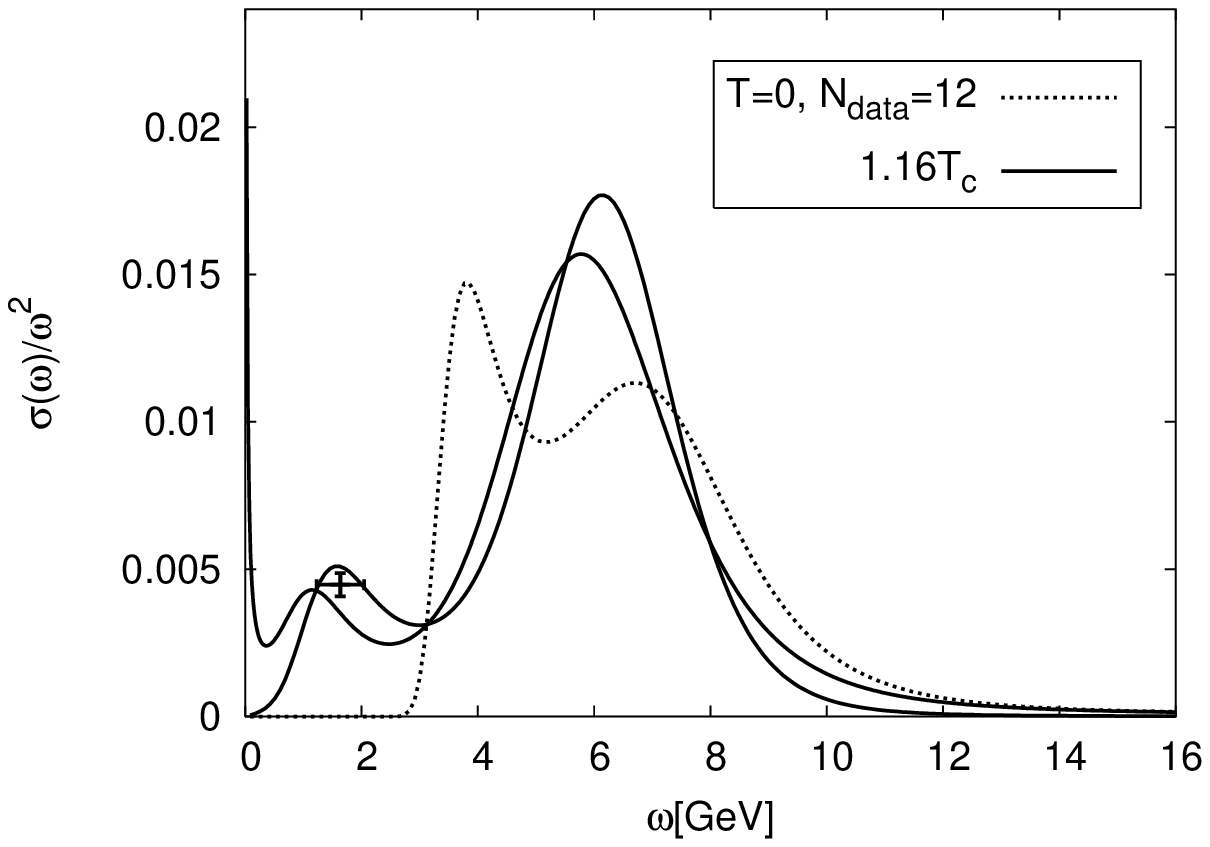}
\vspace*{-1.1cm}
\caption{The charmonium spectral function in the pseudo-scalar (left) and
scalar (right) channels.}
\vspace*{-0.4cm}
\end{figure}
This behavior of the spectral function can be seen already at the level
of the correlation function which does not require invoking statistical tools like MEM \cite{datta02,datta04,jhw}. 
The pseudo-scalar spectral function show almost no temperature dependence across the deconfinement
transition temperature, while the scalar correlator shows large enhancement over the zero temperature correlator
\cite{datta04,jhw}. These features of the correlators are not easy to accommodate in potential model with
screening \cite{mocsy}. 
One of the problems with the spectral functions obtained using MEM is that it shows peaks even at energies
$\omega>5$ GeV where the continuum should dominate. This could be an artifact of the analysis method or the
finite lattice spacing. Also the peaks at low $\omega$ appear to be quite broad (see Fig. 4) and thus may contain 
contributions not only from the ground state but also higher excited states.
Clearly the spectral functions calculated on lattice cannot provide yet quantitative
information about the charmonia properties at finite temperatures. 
The calculations of the charmonium spectral functions 
discussed above were done in quenched approximation. It will be interesting to see how these conclusions are
modified when the effect os dynamical light quarks is taken into account. Preliminary calculation of the charmonium
spectral functions in 2 flavor QCD suggest that the effect of dynamical quarks does not change the above conculsions
significantly \cite{nf2spf}.

There also some preliminary studies of bottomonium correlators and spectral functions \cite{datta06,petrovlat05}. 
An intriguing outcome of these studies is the  enhancement the scalar correlators above deconfinement temperature,
similar to one observed in charmonium case.
The scalar bottomonium correlator corresponds to $\chi_b$ state which is similar in size to 1S charmonia states
and thus is expected to survive above the deconfinement temperature.

Transport processes should appear as peaks in the spectral
functions at very small energies. 
The width of these peaks is related to transport coefficients.
Euclidean correlators are largely insensitive to the detail of these peaks and to very good approximation are
determined only by the area under the peaks \cite{gert,derek}.
Therefore there is little hope of getting transport coefficients from MEM. Nonetheless the presence of a transport contribution
in the vector correlator has been identified \cite{melat05}.

Meson correlators and spectral functions have been calculated also for light quarks \cite{karsch02,asakawa03,gertlat}.
These studies, however, are less systematic than for the charmonium case. One intriguing outcome of these calculations
is the suppression of the vector spectral function at smaller energies which would result in the suppression of the
thermal dilepton and photon rates.  Clearly much more work is needed to verify or rule out this possibility.
The lattice data on the vector correlator on the other hand show only small (about $10\%$ ) deviation from the case of freely
propagating quark anti-quark pair and can provide stringent constraint \cite{karsch02} on the vector spectral 
functions. It would be nice to confront these findings with recent resummed perturbative calculations of the
vector spectral function \cite{guy06}.

At the end of this section let me finally mentions that the temporal correlators and the corresponding
spectral functions are useful not only to study mesonic excitations in high temperature QCD but also any 
real time excitation, including quark and gluon quasi-particles. This, however, requires gauge fixing.
Preliminary results of lattice calculations of quark and gluon
spectral functions in Coulomb gauge have been reported in \cite{melat01}.

\end{document}